\documentclass[a4paper,10pt,twoside]{cpc-hepnp}
\usepackage{fancyhdr}
\usepackage{multicol}
\usepackage{graphicx}
\usepackage{booktabs}
\usepackage{amssymb,bm,mathrsfs,bbm,amscd}
\usepackage{lineno}
\usepackage[tbtags]{amsmath}
\usepackage{lastpage}
\usepackage{subfigure}

\begin{document}


\fancyhead[c]{\small Submitted to Chinese Physics C}



\title{A Study of Energy Correction For the Electron Beam Data in the BGO ECAL of the DAMPE\thanks{Supported by the Chinese 973 Program, Grant No. 2010CB833002, the Strategic Priority Research Program on Space Science of the Chinese Academy of Science, Grant No. XDA04040202-4 and 100 Talents Program of CAS  }}

\author{%
      Zhiying Li$^{1;1)}$\email{LZY0817@mail.ustc.edu.cn}%
\quad Zhiyong Zhang$^{1;2)}$\email{zhzhy@mail.ustc.edu.cn}%
\quad Yifeng Wei$^{1}$
\quad Chi Wang$^{1}$\\
\quad Yunlong Zhang$^{1;3)}$\email{ylzhang@ustc.edu.cn}
\quad Sicheng Wen$^{2}$
\quad Xiaolian Wang$^{1}$
\quad Zizong Xu$^{1}$
\quad Guangshun Huang$^{1;4)}$\email{hgs@ustc.edu.cn}
}
\maketitle

\address{%
$^1$ State Key Laboratory of Particle Detection and Electronics, University of Science and Technology of China, Hefei, Anhui, China, 230026\\
$^2$ Purple Mountain Observatory, CAS,2 West Beijing Road, Nanjing 210008, China \\
}

\begin{abstract}
The DArk Matter Particle Explorer (DAMPE) is an orbital experiment aiming at searching for dark matter indirectly by measuring the spectra of photons, electrons and positrons originating from deep space. The BGO electromagnetic calorimeter is one of the key sub-detectors of the DAMPE, which is designed for high energy measurement with a large dynamic range from 5 GeV to 10 TeV. In this paper, some methods for energy correction are discussed and tried, in order to reconstruct the primary energy of the incident electrons.  Different methods are chosen for the appropriate energy ranges. The results of Geant4 simulation and beam test data (at CERN) are presented.
\end{abstract}

\begin{keyword}
Dark matter, BGO ECAL, energy correction, beam test
\end{keyword}

\begin{pacs}
29.30.Dn , 29.40.Vj , 29.85.+c
\end{pacs}

\footnotetext[0]{\hspace*{-3mm}\raisebox{0.3ex}{$\scriptstyle\copyright$}2013
Chinese Physical Society and the Institute of High Energy Physics
of the Chinese Academy of Sciences and the Institute
of Modern Physics of the Chinese Academy of Sciences and IOP Publishing Ltd}%


\section{Introduction}

The DAMPE is an indirect dark matter searching mission, scheduled to be launched by the end of 2015.  DAMPE detector is designed as search for dark matter indirectly by measuring spectra of photons, electrons along with positrons within a large energy range (from 5 GeV to 10 TeV) and a satisfying energy resolution (1.5$\%$ at 800 GeV) in space.
The DAMPE detector consists of four sub-detectors [3][4], the Plastic Scintillator Detector(PSD), the Silicon Tungsten Trackers (STK), the BGO ECAL and the Neutron Detector (NUD) (Fig.(1)(a)).  The PSD provides heavy ion species discrimination, simultaneously identifies electrons from gamma rays. Tracking information and e/$\gamma$ identification are provided by the STK.  The BGO ECAL is designed as a total absorption detector, measuring the energy deposition due to particle shower and imaging their shower development profile, thus providing a hadron background discriminator.  The NUD, at the bottom of the detector, improves the e/p identification capacity.  The BGO ECAL is composed of 14 layers of BGO crystal bars placed inter-crossly in two dimensions, about 31 radiation lengths in total.  Each layer consists of 22 BGO bars (25 mm*25 mm*600 mm).  There exist 2.5 mm gaps of dead material for support between bars of the same layer and 4 mm gaps between layers (Fig.(1)(b)), and therefore it is necessary to do some corrections in the energy reconstruction of the BGO ECAL.

\begin{figure}
\centering
\subfigure[Structure of DAMPE]{
 \includegraphics[width=6 cm,height=7cm]{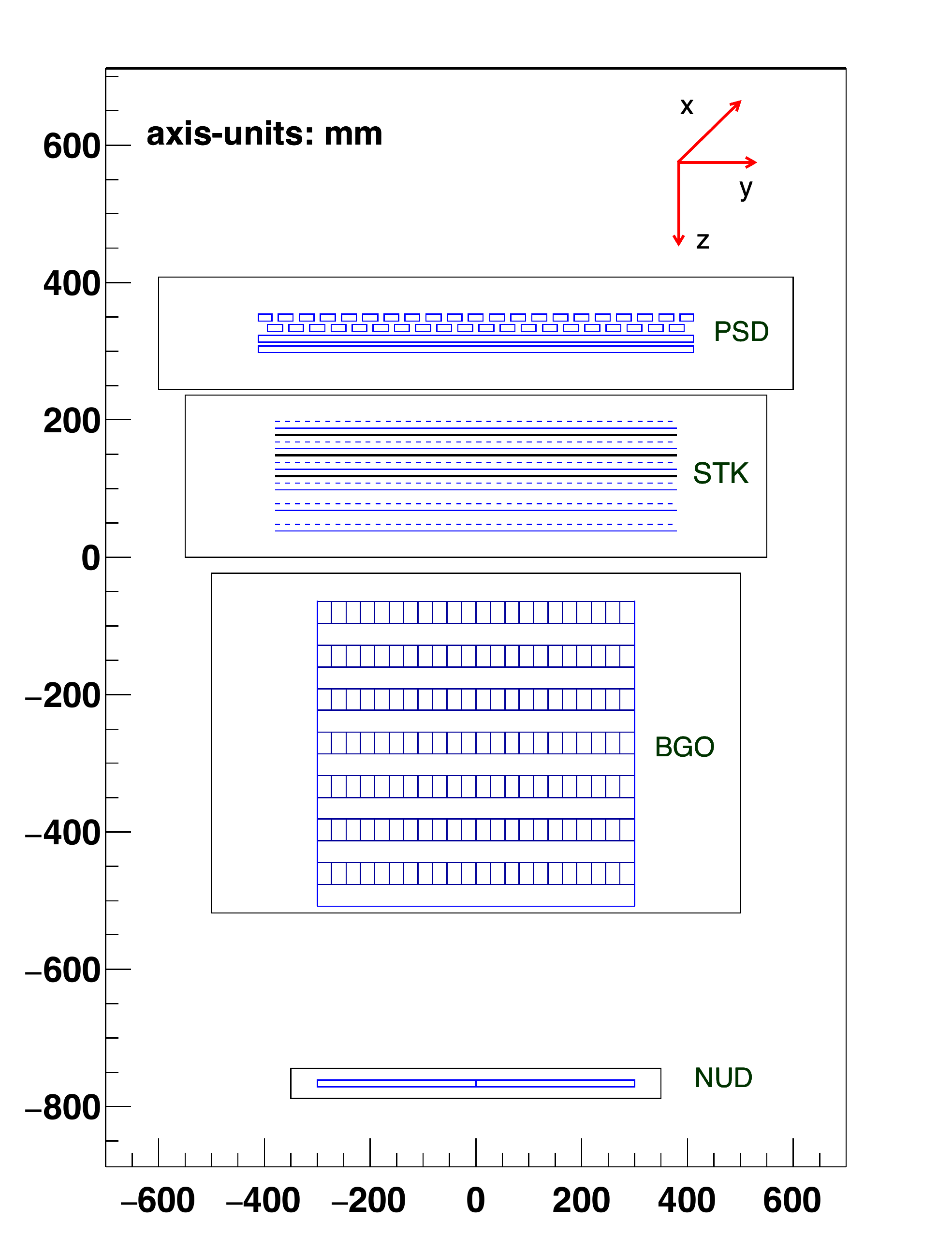}
 }
 \subfigure[Structure of BGO ECAL]{
  \includegraphics[width=8 cm]{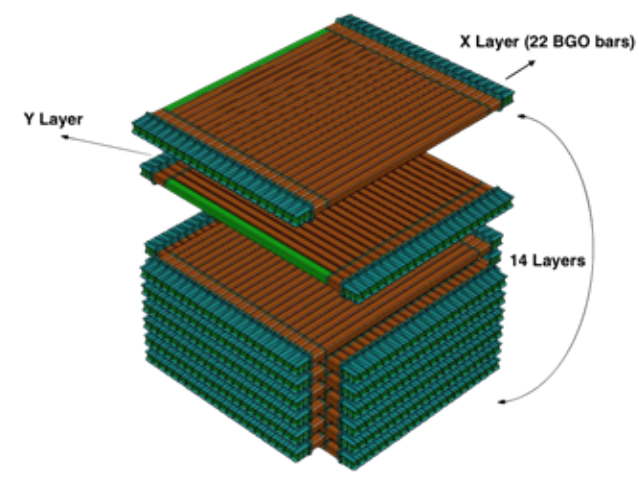}
  }
\figcaption{\label{fig1} Structure of DAMPE detector and BGO ECAL}
  \end{figure} 

\section{Electron Energy Correction of BGO ECAL}

When an electromagnetic cluster in the BGO ECAL is defined, the energy contained in the cluster of calorimeter cells is less than the energy of the incidence electron.  This is due to possible energy losses in dead material between bars of both the same layer and adjacent layers and in front of the calorimeter.  To recover the incidence energy from the measured energy, some correction methods are tried out to find those fit DAMPE most.  The emphasis of this paper is to correct energy of beam test data with electrons below 300 GeV.  We tried the last layer method and similar methods [9], and for $<$300 GeV electron shower is well contained by the 31 radiation-length calorimeter, although for higher energy range around 800 GeV to 10 TeV, backward energy leakage correction would be necessary.  The shower transverse profile method [10] would not give satisfying results due to the granularity of BGO array is not fine enough.  Finally, two main methods, namely S1/S3 and F-Z as described below, are chosen in different energy ranges.  This paper is to present the study on the two energy correction methods [5][6] in two different energy ranges to recover the incident energy of electrons.  The data used for this work is generated from the beam test experiment carried in Nov. 2014 at CERN and electrons are simulated from GEANT4 based programs.  

\subsection{Description of Methods}

  The BGO calorimeter, as mentioned above, has dead materials between crystal bars.  As shown in Fig. 2, for electrons passing the calorimeter through different positions, the simulation data (Fig. 3) shows that energy deposits are quite different with electrons injecting positions.
 
 \begin{center}
\includegraphics[width=10cm]{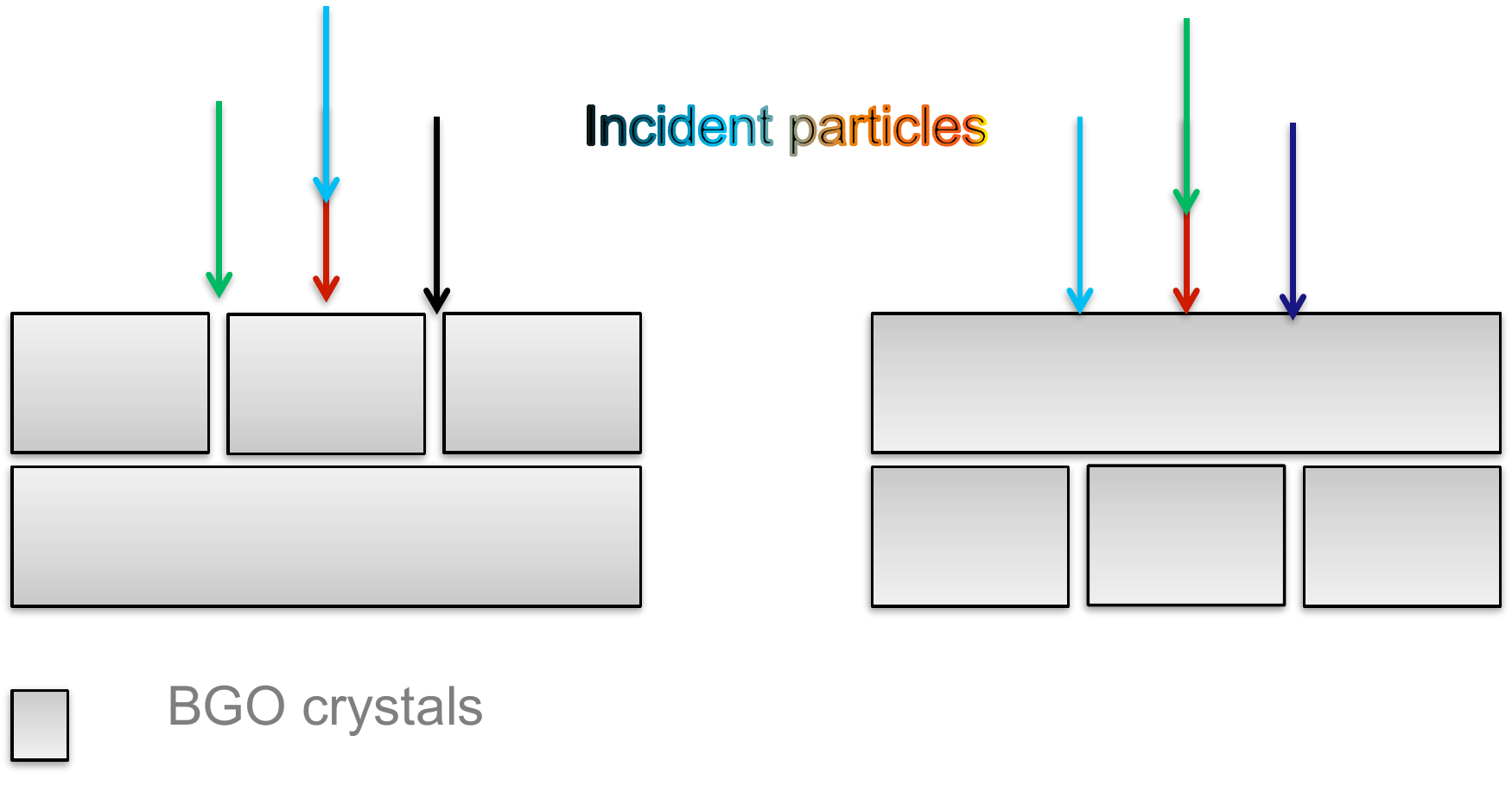}
\figcaption{\label{fig2} Possible Incidence Electron Positions}
\end{center}

\begin{center}
\includegraphics[width=10cm]{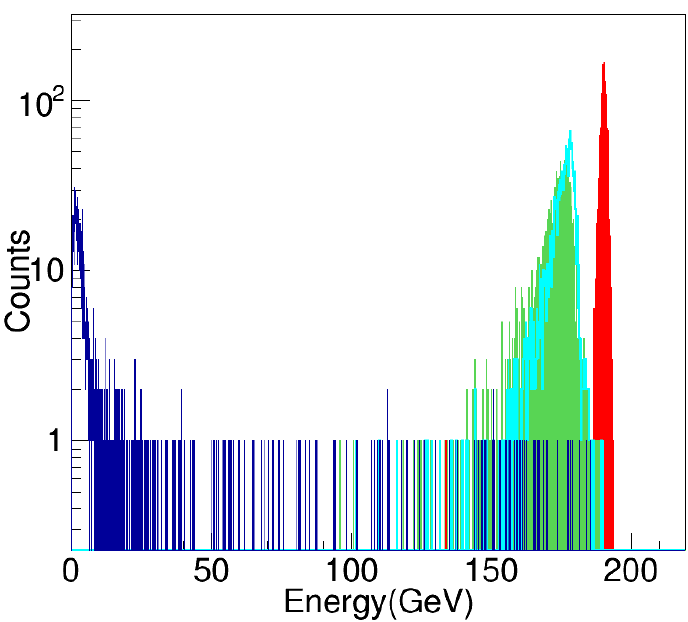}
\figcaption{\label{fig3} The distribution of energy deposit of 200 GeV electron beam source with a spot size of a few centimeters(MC data)}
\end{center}

\subsubsection{S1/S3 Method}

The S1/S3 method [5][6] takes into account small dead material between neighboring cells, especially cells of the same layer. For layer i, the maximum energy deposited in a bar is s1[i], and sum of s1[i] and its 2 adjacent bars is s3[i], then the S1/S3 ratio is defined as  Equ.(1). 
\begin{eqnarray}
\label{eq1}
ratio = \frac{S1}{S3} = \frac{\sum_{i} s1[i]}{\sum_{i} s3[i]}             
\end{eqnarray}

The effect of dead material between cells appears to be related to the ratio between S1 and S3.  As for DAMPE simulated 5 GeV electrons (Fig. 4), the ratio is about 0.5 for particles hitting the calorimeter in the dead space, in connection with most energy loss, and about 0.9 for particles impinging at bars' centers , with energy loss.  F is defined as ratio between deposit energy in BGO ECAL and incidence energy of electrons.  The S1/S3 ratio versus F relation of electrons generated by simulation based on GEANT4 is given in Fig. 4, which shows it can be divided into three parts, fitted by a constant function and two linear functions as given in Equ.(2)). 

\begin{eqnarray}
\label{eq2}
F(x) = \begin{cases}
p_{0}&\text{x  $< p_{1}$},\\[1mm]
p_{1}x + p_{0} - p_{2}p_{1}&\text{$p_{1} \leq$x $< p_{3}$},\\[1mm]
p_{4}x + p_{2}p_{3} + p_{0} - p_{2}p_{1} - p_{3}p_{4}&\text{x $\geq p_{3}$}   
\end{cases}
\end{eqnarray}

\begin{center}
\includegraphics[width=12cm]{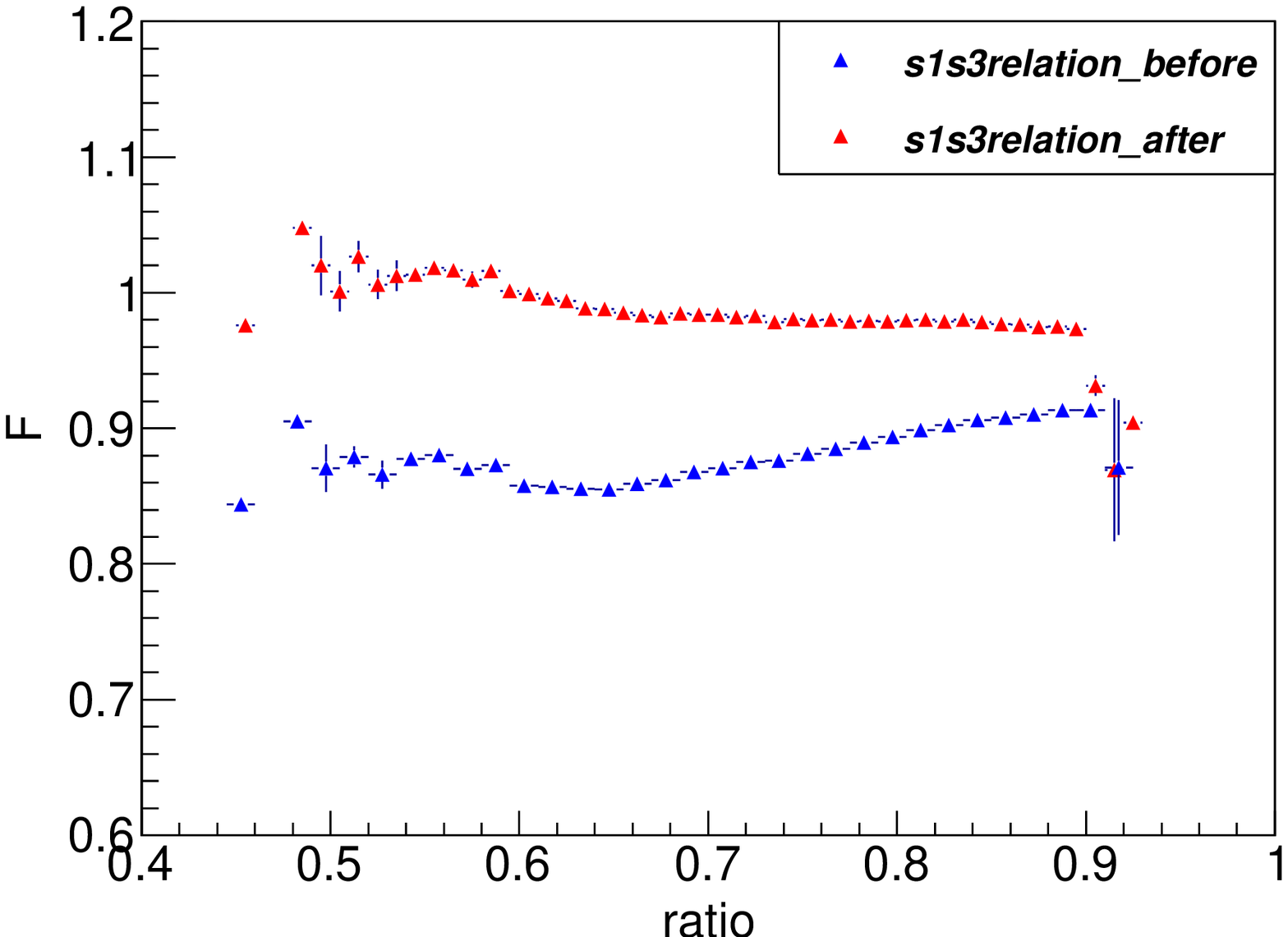}
\figcaption{\label{fig4} Ratio-F relation before and after correction of 5 GeV electrons generated by simulation}
\end{center}

The first two parts are used in the correction.  The features of fit parameters will be discussed in the following sections.  As is given in Fig. 4, the F value after correction is much close to 1.0, while before correction it could vary from 0.84 to about 0.92.  The effect of S1/S3 correction method for 5GeV electrons is given in Fig. 5 and it can be seen that mean of energy distribution before correction is around 4.5 GeV while after S1/S3 correction it becomes close to 5 GeV, which is the incidence energy.

\begin{center}
\includegraphics[width=12cm]{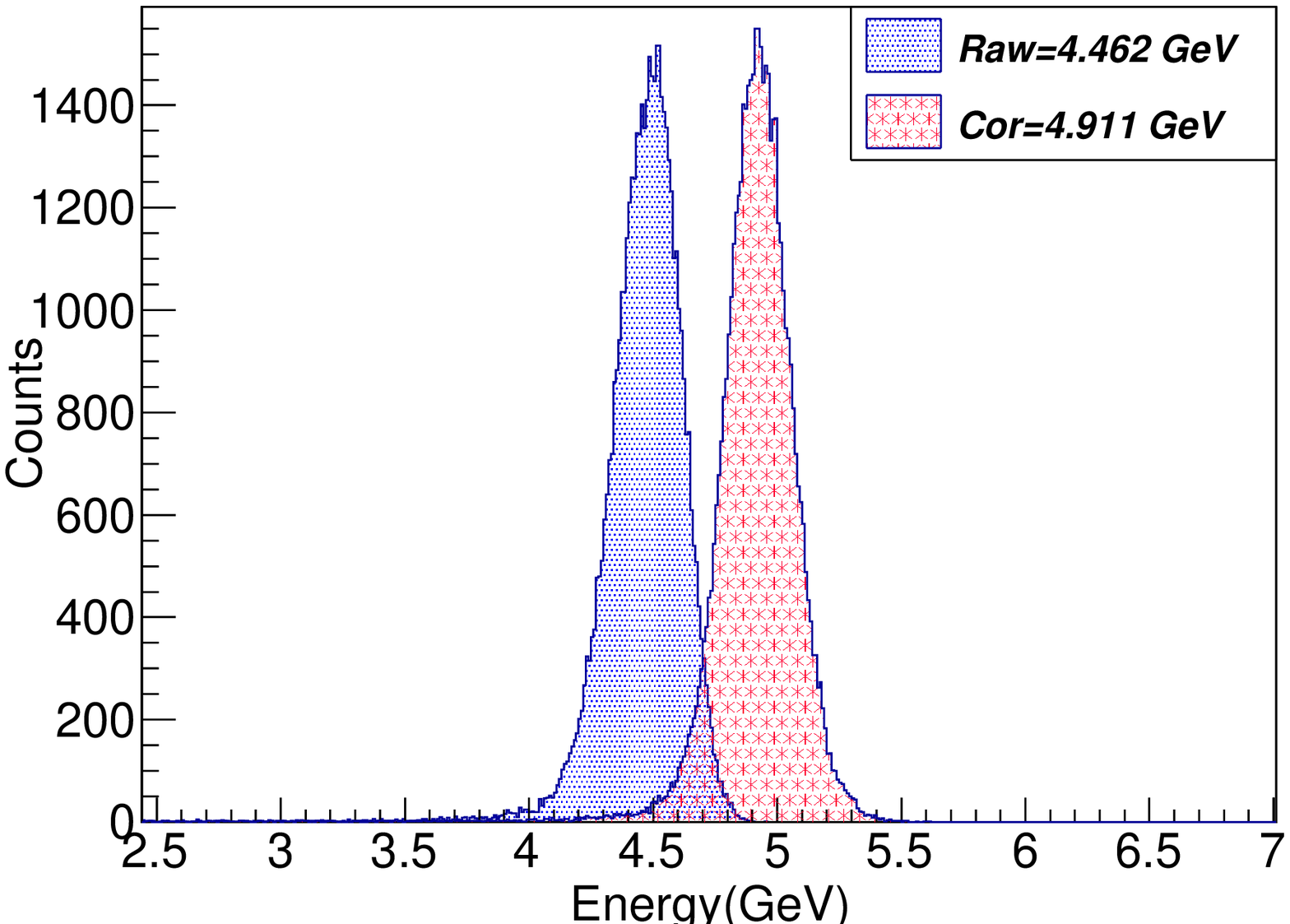}
\figcaption{\label{fig5} Energy distribution before and after S1/S3 correction of simulated 5 GeV electrons, Raw for raw data energy and Cor for corrected energy}
\end{center}

\subsubsection{F-Z Method}

The F-Z method [7] takes into accounts both gaps between BGO bars of the same layer and gaps between the layers.  F-Z method uses parameters F, which is defined as the ratio of energy given by the calorimeter against incident energy, and Z$_{bary}$, defined as the shower development longitude barycenter of incident electrons (defined as Equ.(3)). 
\begin{eqnarray}
\label{eq3}
Z_{bary} = \frac{\sum_{i} E_{i}Z_{i}}{E_{tot}}    
\end{eqnarray}
E$_{i}$ is defined as the energy deposit in the Z$_{i}$-layer.   Z value is measured by radiation length of BGO crystal. The F-Z relation of 50 GeV electrons provided by GEANT4 based simulation is shown in Fig. 6.

\begin{center}
\includegraphics[width=12cm]{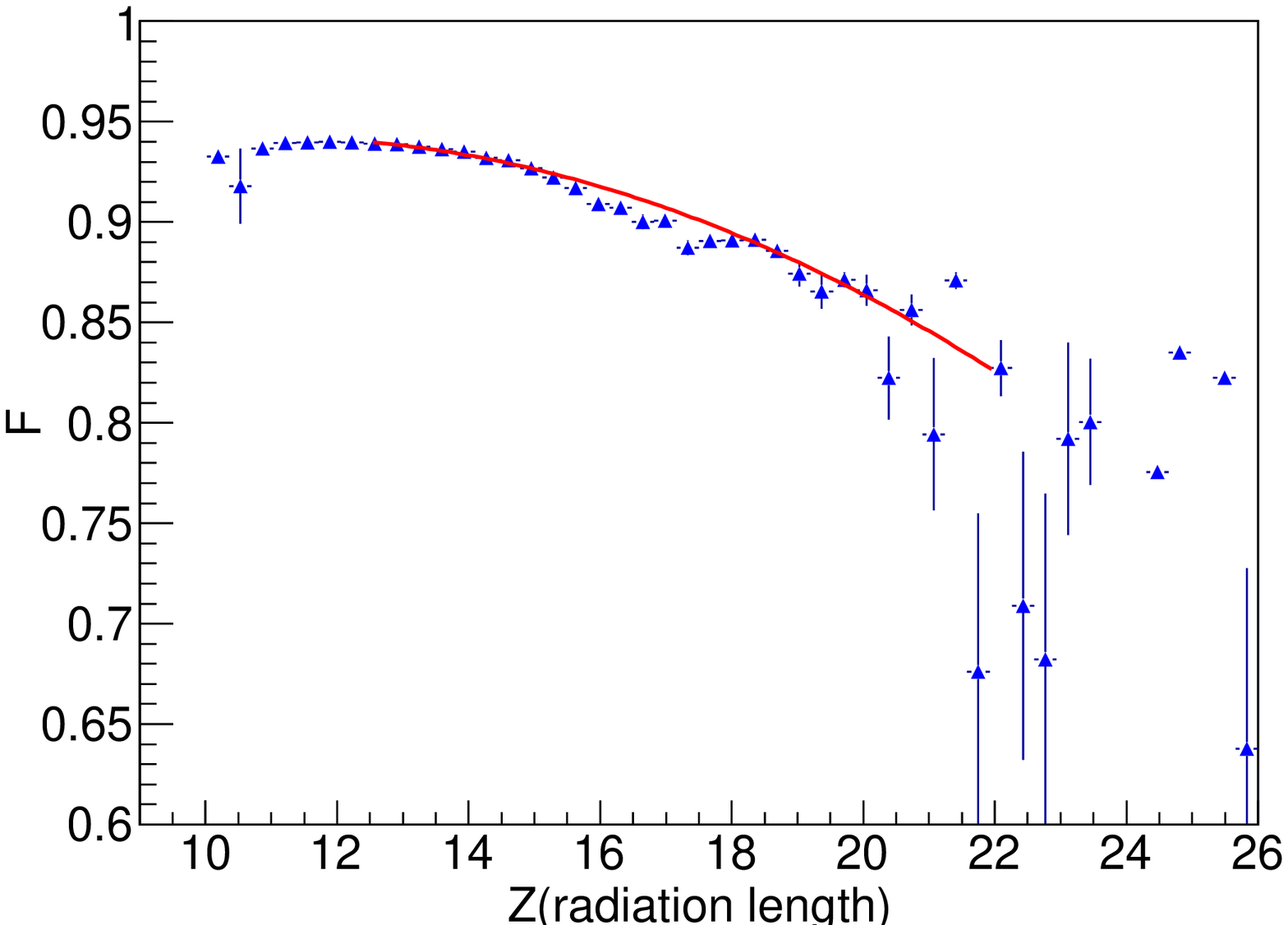}
\figcaption{\label{fig6} F-Z relation of 50GeV electron simulation data}
\end{center}

The F-Z relation indicates that with the Z value getting larger, the ratio of energy given by BGO calorimeter against the incident energy of electrons would become smaller. The Z value getting bigger suggests the electron goes through more BGO crystal layers, which could cause more energy loss in gaps between the layers.  The F-Z relation is fitted by a quadratic Equ.(4)) (the fitting curve is shown as the red line in Fig. 6). With the F-Z relation, the energy given by the calorimeter is corrected to the incident energy.  

\begin{eqnarray}
\label{eq4}
F(z) = p_{0} + p_{1}z + p_{2}z^2   (4)
\end{eqnarray}

\subsection{Choice of methods}

The suitable method within different energy range is chosen according to its energy correction performances.  The S1/S3 method is not applied to higher energies, as the fit function could not describe the relation well (as shown in Fig. 7).  The ratio-F relation for 50 GeV electrons is more curved than linear so that the fit function used for S1/S3 correction would no longer be suitable.  In this way, for energies above 20 GeV, S1/S3 correction method is replaced by F-Z method.  The parameters used here and in following energy correction for beam test results are generated from beam test data results.
\begin{center}
\includegraphics[width=12cm]{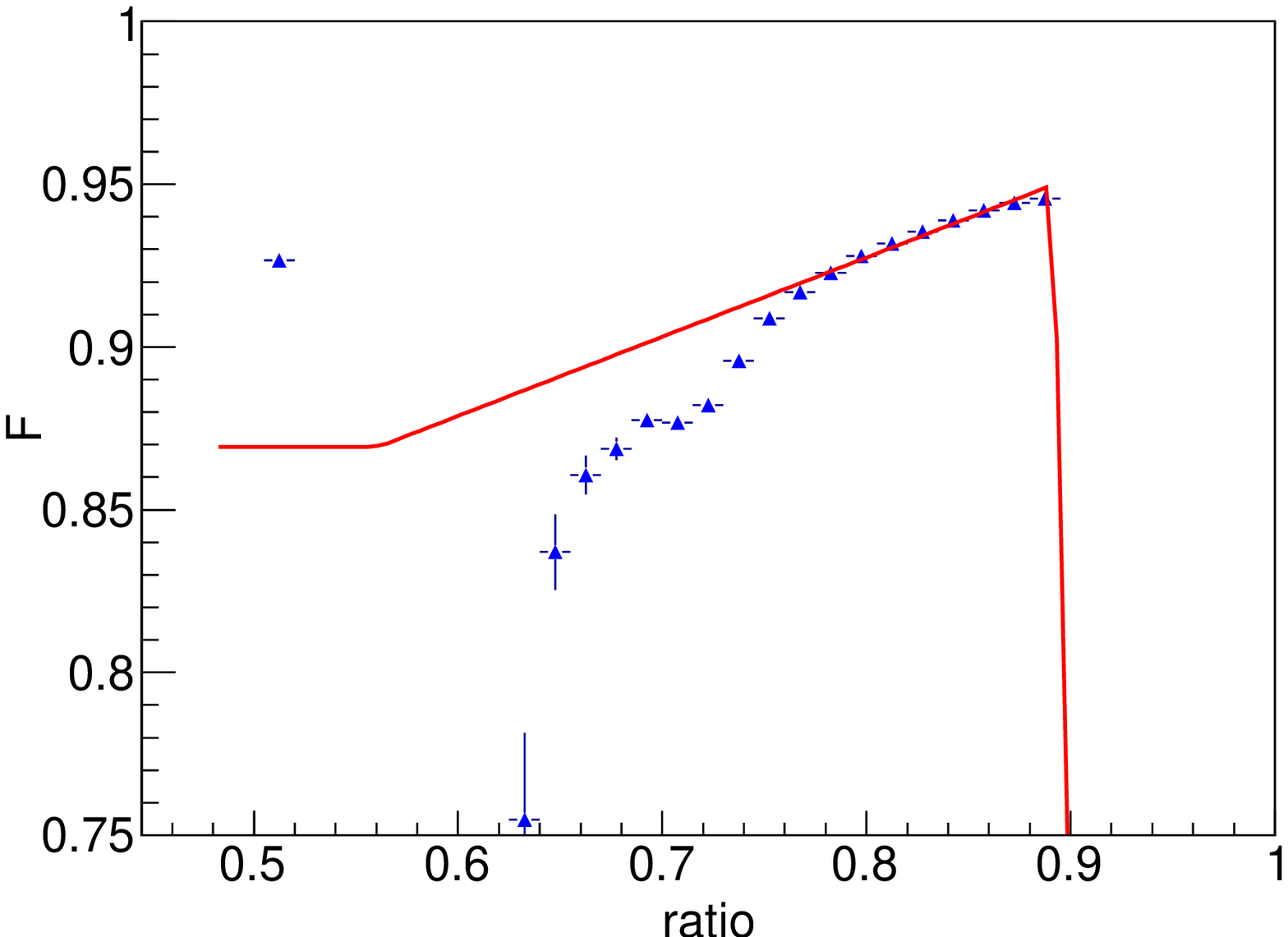}
\figcaption{\label{fig7} S1/S3 ratio-F relation of 50 GeV electrons(MC data, red line from fit function Equ. (2))}
\end{center}

Due to the requirements of the correction, we decide to use S1/S3 method within the energy range of 3 GeV to 20 GeV and use F-Z method in the range of 20 GeV up to 300 GeV.   We studied the parameters of both methods and the results are given in following parts.

\subsection{Parameter-energy Relation}

\subsubsection{S1/S3 parameter-energy relation}

For study of the energy dependence of fit parameters, the MC data samples with various electron energy range from 3 GeV to 300 GeV are produced and reconstructed; the plots like Fig.4 and Fig.6 are obtained and parameter fits are made with Equ.(2) and Equ.(4) respectively.  From the Fig.8, the fit parameters in Equ.(2) of S1/S3 method do not change much in concern of energy changes.  This result indicates that for different energy points within the energy range of 3 GeV to 20 GeV, it is satisfying to use the same set of parameters for S1/S3 method.
\begin{center}
\includegraphics[width=16cm]{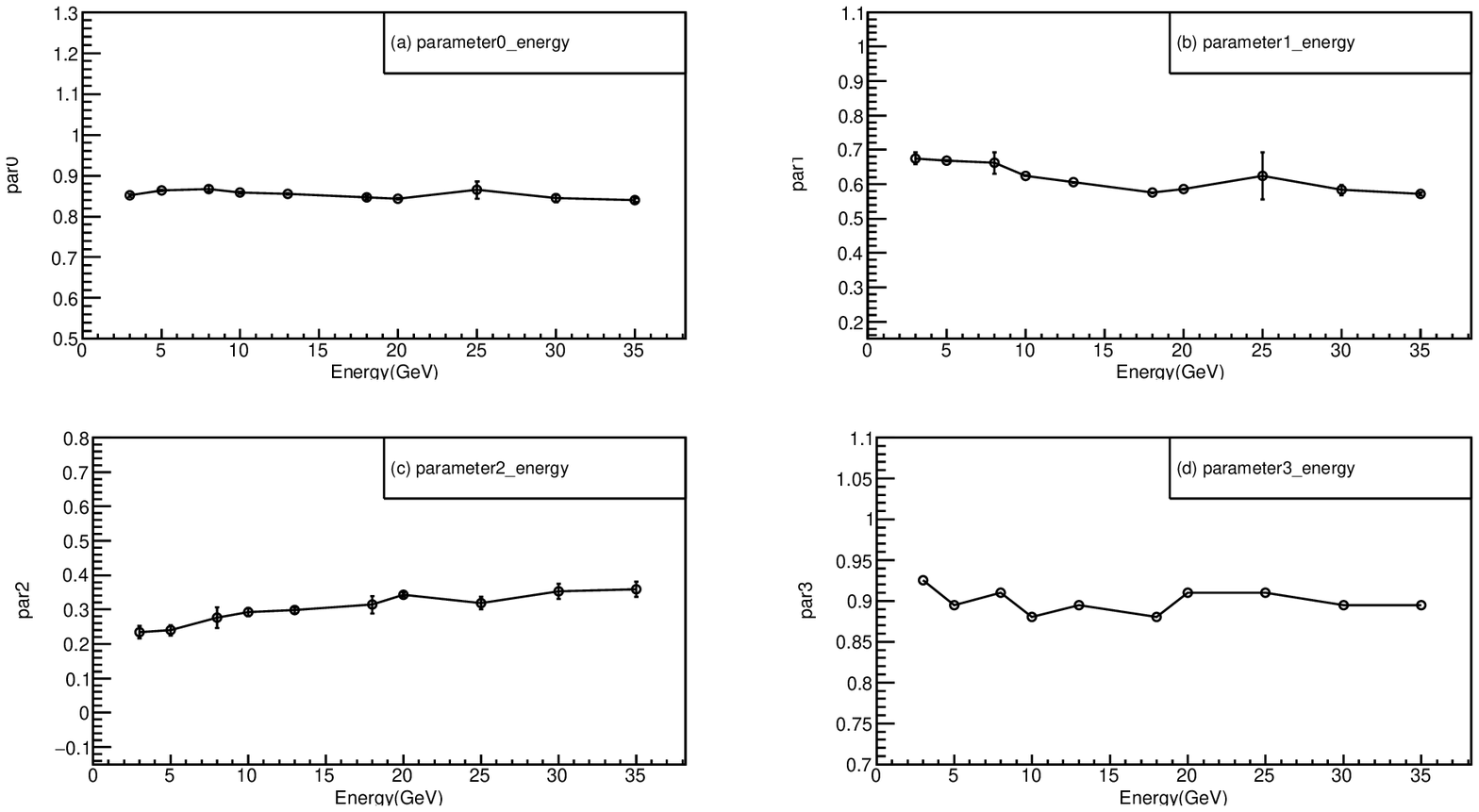}
\figcaption{\label{fig8} Parameter-energy relation of S1/S3 method}
\end{center}

\subsubsection{F-Z method parameter-energy relation}

The relation between parameters in Euq.(4) of the fit function of F-Z method and energy given by the calorimeter is also studied with the Monte Carlo data sample mentioned above, Fig. 9 shows there are no significant energy dependence of fit parameters in Equ.(4).As is shown in Fig. 9, it is viable to use the same set of parameters throughout the energy range of 50 GeV to 300 GeV for F-Z method.

\begin{center}
\includegraphics[width=16cm]{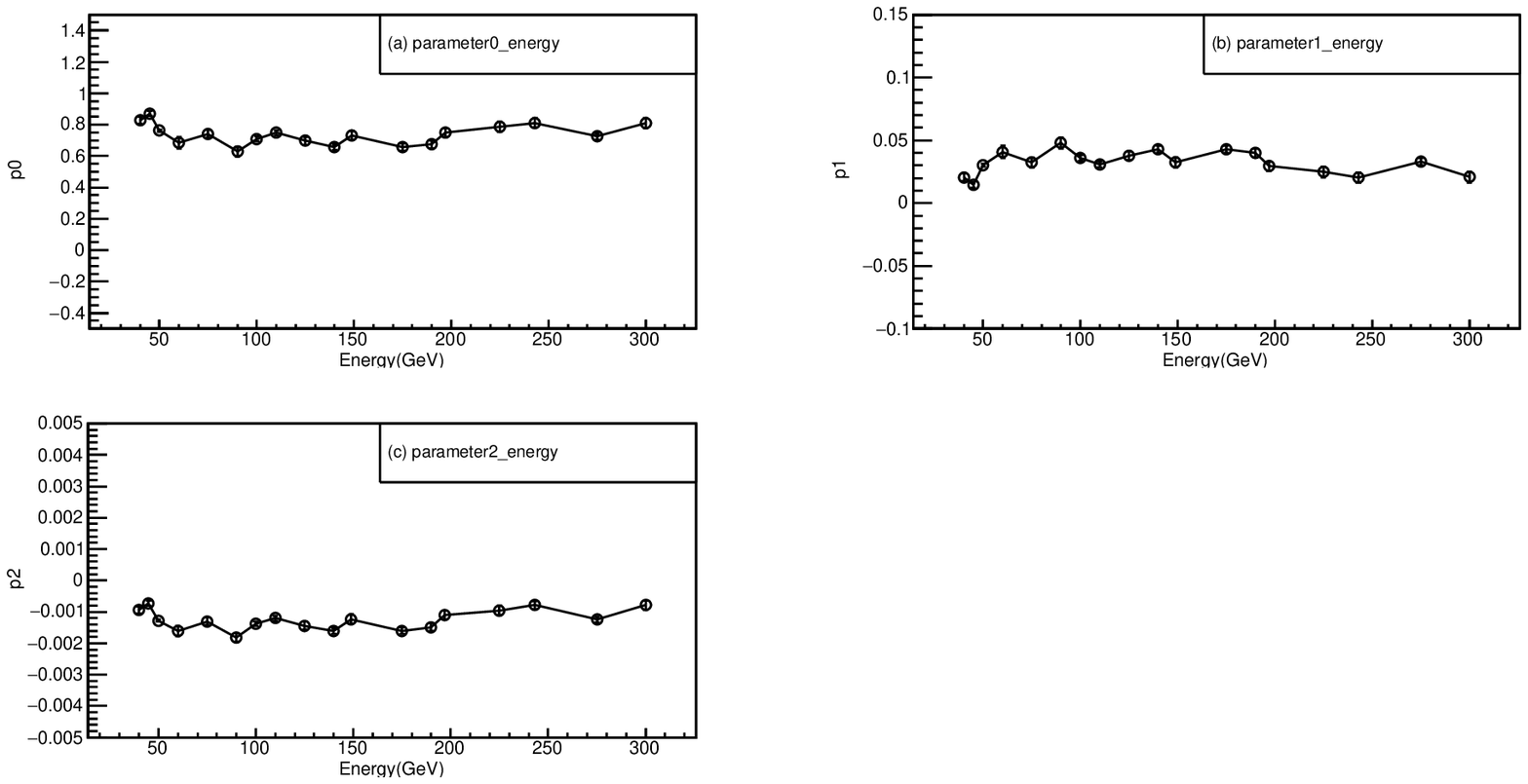}
\figcaption{\label{fig9} Parameter-Energy relation of F-Z method, electrons with incidence angle of $0^{\circ}$}
\end{center}
\subsection{Parameter-Incidence Angle Relation}

The beam test also used electrons with different incidence angles ranging from 0 to 30 degree, and studies have been carried on to find out the relation between incidence angles and fit parameters with GEANT4 based simulation, which can be seen in Fig. 10 (50GeV simulated electrons).  Within the small degrees up to 30 degree as is concerned in beam tests, the fit parameters do not respond to incident angle changes.  As far as is concerned in beam test data, there exists no transverse leakage caused by incidence angles.  Since beam test data with incidence angles are only available for energy above 50 GeV, parameter-incidence angle relation is studied for F-Z method alone.
\begin{center}
\includegraphics[width=16cm]{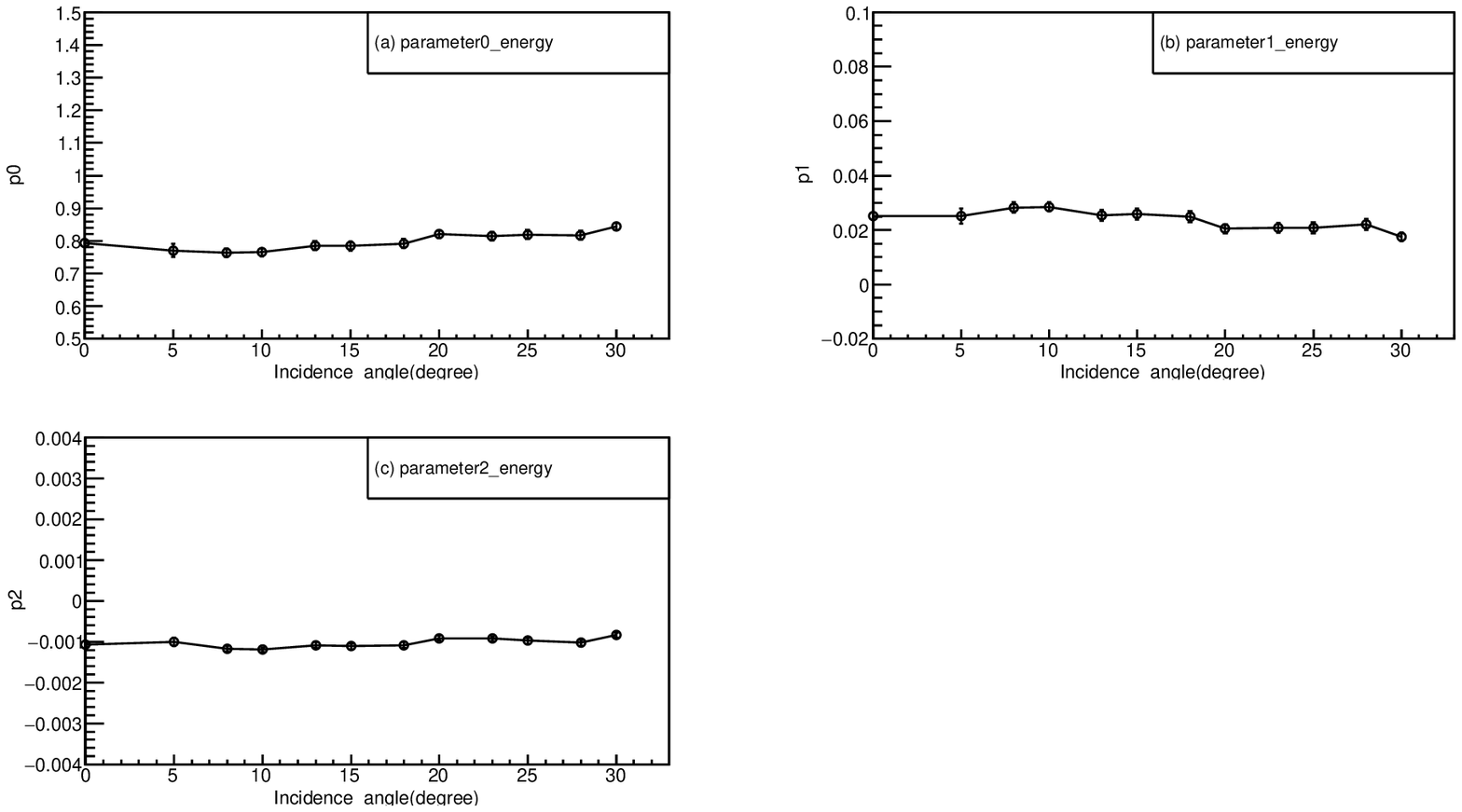}
\figcaption{\label{fig10} Parameter-Incidence Angle relation of 50 GeV simulated electrons}
\end{center}

To study the influence of electron energies with a given incidence angle, more simulation is done.  No evidence indicates that fitting parameters of electrons with incidence angles are energy dependent (seen from Fig.11).
\begin{center}
\includegraphics[width=16cm]{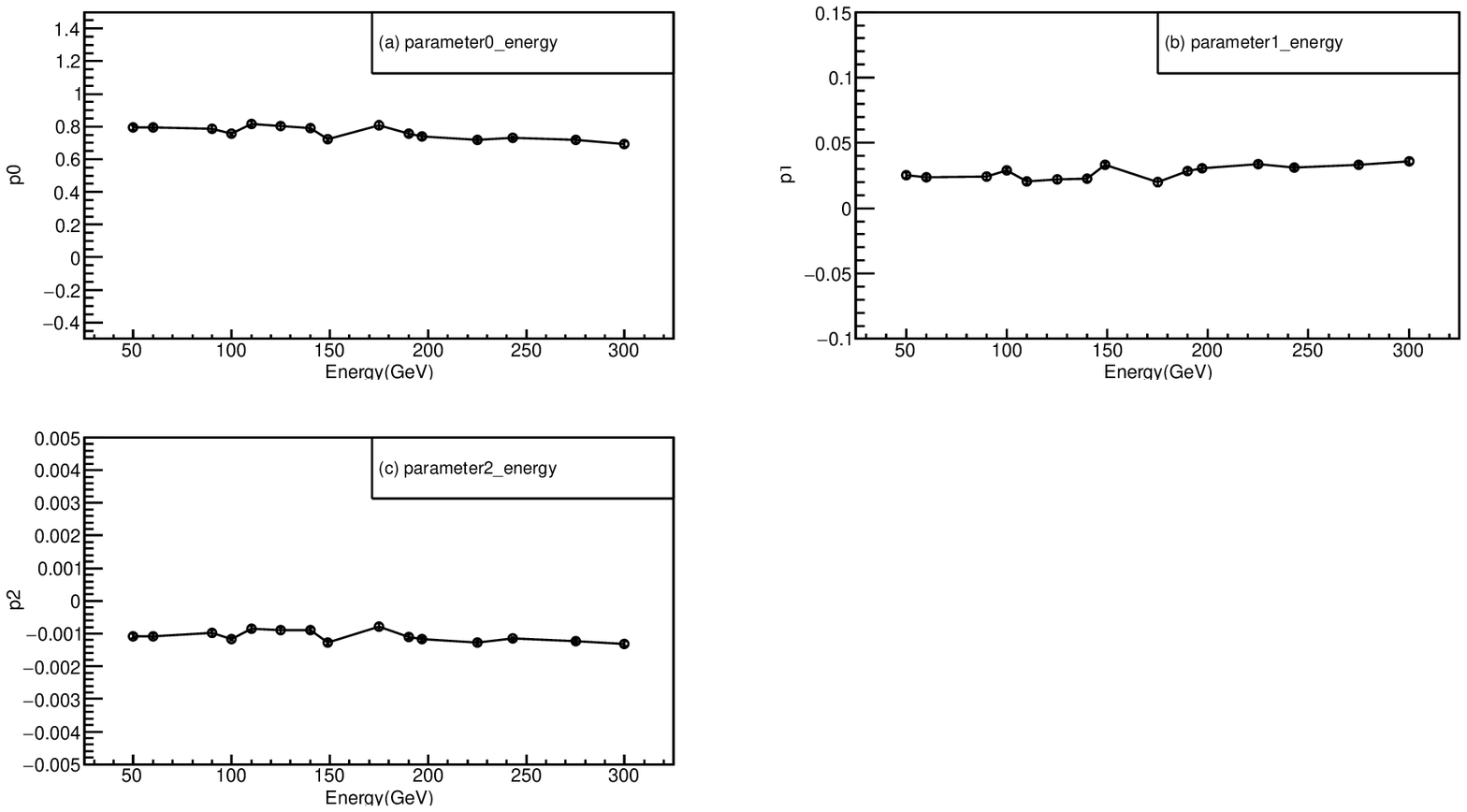}
\figcaption{\label{fig11} Parameter-Energy relation of simulated electrons with the incidence angle of $15^{\circ}$}
\end{center}

\section{Correction Results of Beam Test Data}

The energy correction results for electron beam test data using S1/S3 method at GeV (Figure 12(a)) and results using F-Z method at 50 GeV(b), 100 GeV(c), 149 GeV(d), 197 GeV(e) and 243 GeV(f) are shown in Fig.12 as examples.  Energy correction results of 50GeV electrons from beam test with different incidence angles are given in Fig 13.  Table.1 tabulates the ratios of the reconstructed energies and the energy resolution before and after correction and Fig.14 shows the ratio of the corrected measured energies and to incidence energies (left) and energy resolution before(green dots) and after correction.
\begin{center}
\includegraphics[width=20cm]{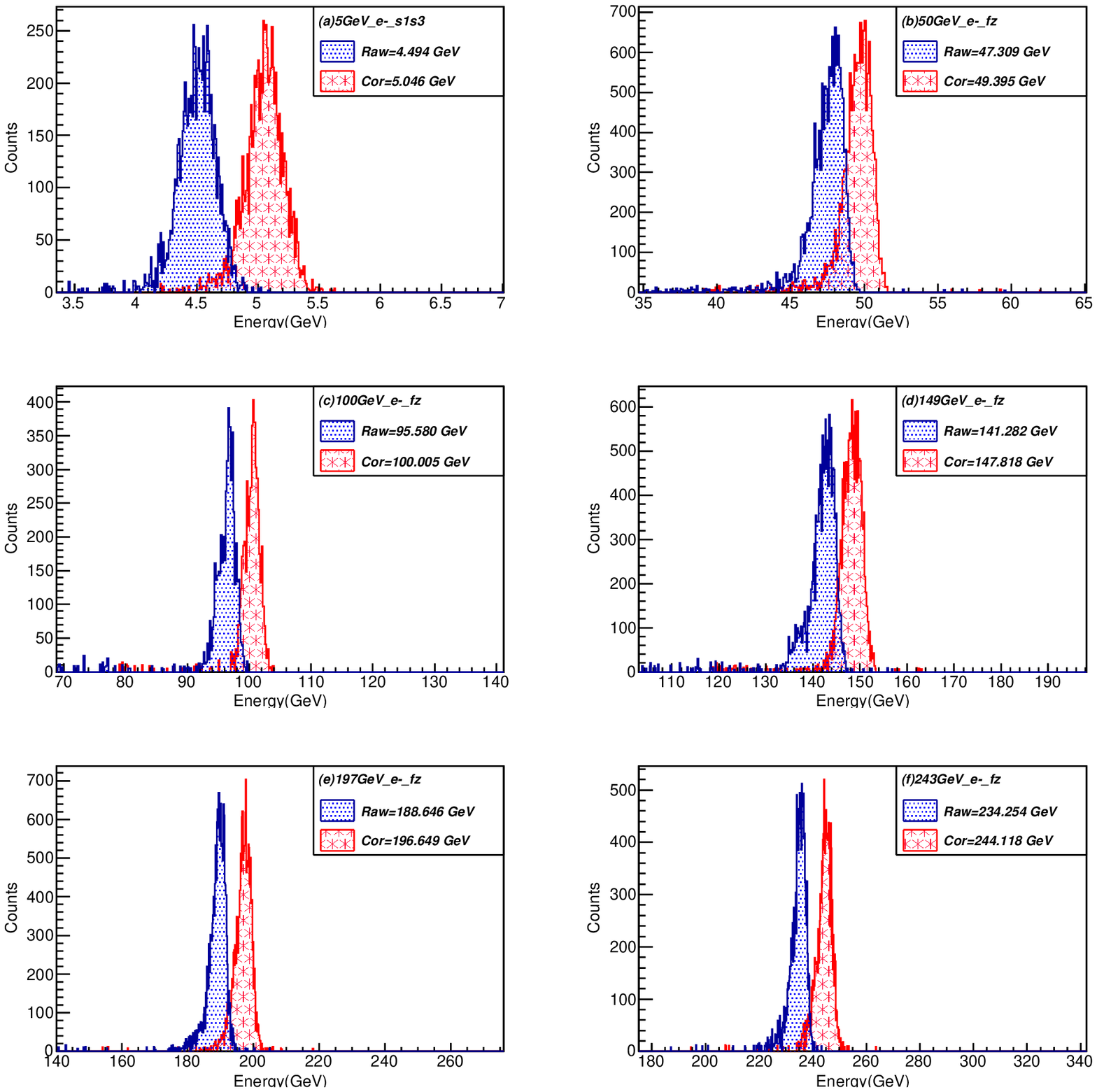}
\figcaption{\label{fig12} Beam Test electrons energy distribution before and after correction, Raw for mean energy for raw data and Cor for mean energy after correction}
\end{center}
\begin{center}
\includegraphics[width=20cm]{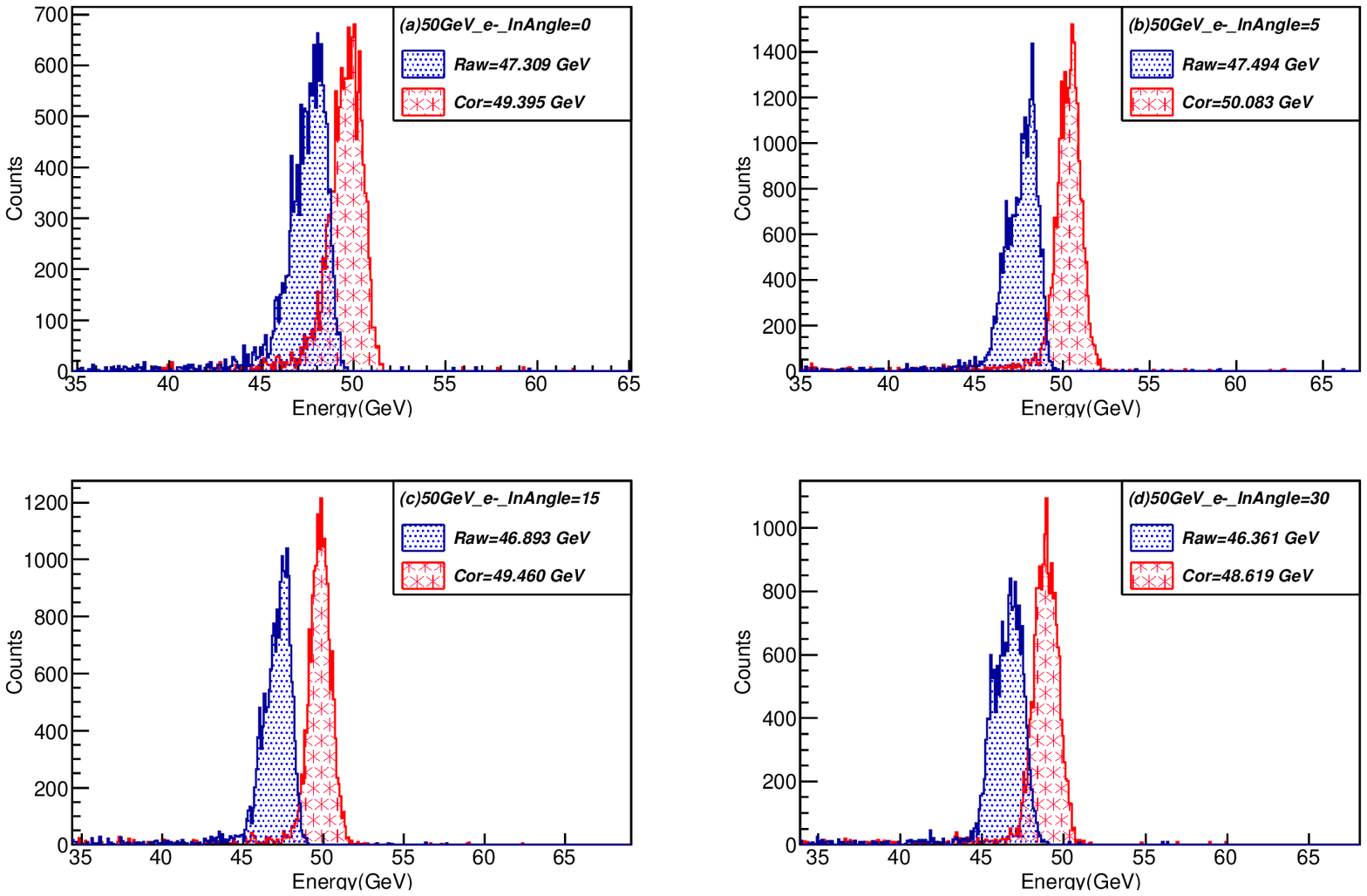}
\figcaption{\label{fig13} Beam Test 50 GeV electrons with various incidence angles energy distribution before and after correction}
\end{center}

\begin{center}
\tabcaption{\label{tab1} Energy ratio and energy resolution before and after correction }
\footnotesize
\begin{tabular*}{170mm}{@{\extracolsep{\fill}}ccccccc}
\toprule InEnergy/GeV & Raw/GeV & $Corrected$/GeV & $Raw Ratio$/$\%$ & $Corrected Ratio$/$\%$ & $Raw Resolution$/$\%$ & $Corrected Resolution$/$\%$
\\
\hline
3& 2.643 & 2.951 & 88.1 & 98.4 & 4.26 & 3.64 \\
4 & 3.568 & 3.973 & 89.2 & 99.33 & 3.38 & 3.13 \\
5 & 4.503 &5.009 & 90.06 & 100.18 & 2.96 & 2.64 \\
50 & 47.57 & 49.64 & 95.14 & 99.28 &1.72 & 1.5 \\
100 & 96.42 & 100.5 & 96.42 & 100.5 & 1.36 & 1.14 \\
149 & 142.5 & 148.3 & 95.64 & 99.5 & 1.41 & 1.29 \\
197 & 189.2 & 197 & 96.04 & 100 &1.05 & 1.09 \\
243 & 235 & 244.7 & 96.71 & 100.6 & 1.00 &0.99 \\
\bottomrule
\end{tabular*}%
\end{center}

\begin{center}
\includegraphics[width=20cm]{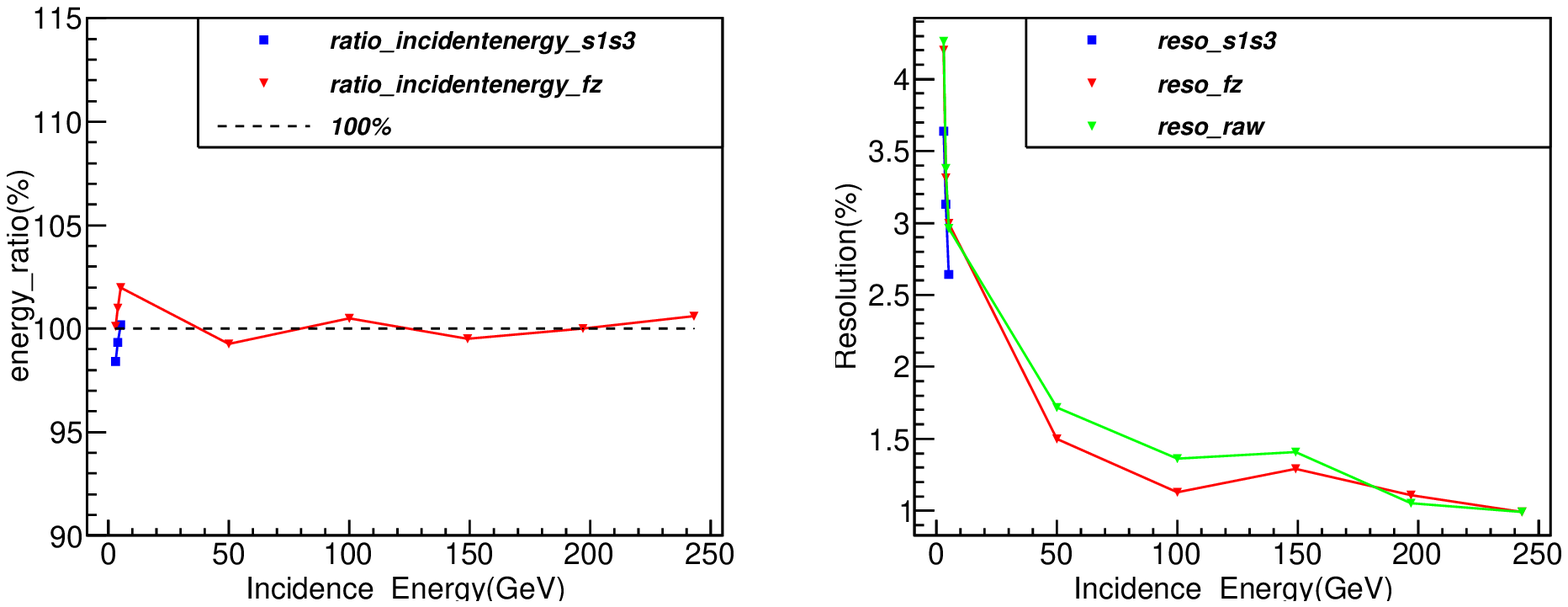}
\figcaption{\label{fig14} Energy-reconstructed versus incidence energy and energy resolution versus incidence energy, energy$\_$ratio is defined as ratio corrected energy against incidence energy}
\end{center}

\section{Conclusion}

Two correction methods are applied to the data samples both from simulation and beam test with electron energies from 3 to 243 GeV, achieving corrected energy versus incidence energy ratio ~100$\%$.  S1/S3 method is applied within the energy range of 3 GeV to 20 GeV and F-Z method is applied within the energy range of 50 GeV to 243 GeV.  The systematic uncertainties of energy measurement have been improved.   The energy resolutions are a little better after corrections.  The corrected energy resolution is below 4$\%$ for electrons under 10 GeV and about 1$\%$ for electrons above 100 GeV.

\acknowledgments{This work was supported by the Chinese 973 Program, Grant No. 2010CB833002, the Strategic Priority Research Program on Space Science of the Chinese Academy of Science, Grant No. XDA04040202-4 and 100 Talents Program of CAS. We would like to thank colleagues of DAMPE, colleagues of University of Geneva and University of Lecce, University of Bari, and CERN test beam team. }


\vspace{10mm}

\vspace{-1mm}
\centerline{\rule{80mm}{0.1pt}}
\vspace{2mm}

\begin{multicols}{2}

\end{multicols}

\clearpage

\end{document}